\documentclass[12pt]{iopart}
\usepackage{iopams}  
\usepackage{epsfig}
\usepackage{graphicx}
\usepackage{epstopdf}

\def\slashchar#1{\setbox0=\hbox{$#1$}     		
   \dimen0=\wd0                                 	
   \setbox1=\hbox{/} \dimen1=\wd1               	
   \ifdim\dimen0>\dimen1                        	
      \rlap{\hbox to \dimen0{\hfil/\hfil}}      	
      #1                                        	
   \else                                        	
      \rlap{\hbox to \dimen1{\hfil$#1$\hfil}}   	
      /                                         	
   \fi}   
\def\bs{\boldsymbol}
\def\wt#1{\widetilde{#1}}
\def\p{{\boldsymbol p}}

\def\x{{\boldsymbol x}}
\def\y{{\boldsymbol y}}

\def\u{{\boldsymbol u}}
\begin{document}

\title{From Glasma to Quark Gluon Plasma in heavy ion collisions
}

\author{Raju Venugopalan}

\address{Department of Physics,
Bldg.\ 510A, 
Brookhaven National Laboratory,\\
Upton NY 11973, USA}
\ead{raju@bnl.gov}

\begin{abstract}
When two sheets of Color Glass Condensate collide in a high energy heavy ion collision, they form matter with very high energy densities called the Glasma. We describe how this matter is formed, its remarkable properties and its relevance for understanding thermalization of the  
Quark Gluon Plasma in heavy ion collisions. Long range rapidity correlations contained in the near side ridge measured in heavy ion collisions may allow one to directly infer the properties of the Glasma. 
  
\end{abstract}

\section{Introduction}

At very high energies, multi-particle production in QCD is generated by low $x$ partons in the nuclear wavefunctions. These partons 
have properties best described as a Color Glass Condensate (CGC)~\cite{CGC}. When  two sheets of  CGC collide in a high energy heavy ion collision, these partons are released and create energy densities an order of magnitude above the energy density required for the crossover from hadronic to partonic degrees of freedom. This matter, at early times after a heavy ion collision, is a coherent classical field, which expands, decays into nearly on shell partons and may eventually thermalize to form a Quark Gluon Plasma (QGP).  Because it is formed by melting the frozen CGC degees of freedom, and because it is the non-equilibrium matter preceding the QGP, this matter is called the Glasma~\cite{Glasma}. 

The Glasma is of intrinsic interest because  heavy ion collisions create  chromo-electric and magnetic fields in bulk, that  in absolute magnitude, are some of the strongest such fields in nature; they are greater (by several orders of magnitude) than the magnetic fields on the surface of magnetars. The Glasma is also 
important for quantifying the initial conditions for the QGP. In particular, two measures of QGP formation, the flow of bulk matter and the energy loss of jets, can be significantly influenced by the properties of the Glasma. 
In the former case, the initial matter distribution and non-equilibrium flow are important for a quantitative determination of the properties of the ``perfect" fluid QCD. In case of the latter, jets can experience significant energy loss in their interactions with the Glasma. 

There is a strong analogy between the physics of the little bang in a heavy ion collision and the big bang that created our universe. In the big bang, the inflaton field with large occupation number $O(\frac{1}{g^2})$ 
decays rapidly with the expansion of the universe. Likewise, in the little bang, the Glasma field with occupation number $O(\frac{1}{g^2})$ decays rapidly after the collision. In the big bang, low momentum quantum fluctuations are explosively amplified in a process known as pre-heating~\cite{Bellido}. In the little bang, the explosive amplification of low momentum quantum fluctuations may be related to a Weibel~\cite{Stan} or a 
Nielsen-Olesen~\cite{N-O} type instability. The interaction of quantum fluctuations with the decaying inflaton field can lead to rapid ``turbulent" thermalization~\cite{Tkachev}. A similar phenomenon may be 
responsible for rapid thermalization in heavy ion collisions~\cite{turbulence}. There is also likely a concrete analogy between super horizon fluctuations 
observed in the COBE and WMAP measurements and the near side Ridge measured at RHIC~\cite{Ridge-expt}. Another strong analogy is between sphaleron driven topological transitions that may induce $P$ and $CP$ odd metastable states in heavy ion collisions~\cite{Warringa} and the matter--anti-matter asymmetry generated by $C$ and $CP$ violating topological transitions during electroweak baryogenesis. What both long range rapidity correlations and topological transitions in strong fields have in common is the likelihood that they both survive the later stage interactions that  
thermalize the system causing it to lose memory of how it was formed. 

In this talk, I will first briefly describe our understanding of the properties of hadron and nuclear wavefunctions at high energies in the CGC framework. I will next describe  the 
quantum field theory framework of particle production in strong time dependent fields. Multi-particle production in the Glasma can be computed systematically in this framework.  An important part of this 
systematic computation is a proof of high energy factorization. I will describe briefly how plasma instabilities arise and are accounted for in this framework. Finally, I will discuss how Glasma flux tubes form the near side ridge seen in heavy ion collisions.

\section{Before the little bang}
\label{sect:two}
At high energies, the competition between QCD bremsstrahlung which enhances the parton density)and screening/recombination processes which depletes it leads to a saturation of parton densities when the field strengths squared become maximal: $E^2\sim B^2 \sim {1\over \alpha_S}$. This non-linear strong field regime of QCD is  characterized by a saturation scale~\cite{GLR} $Q_S(x,A)$; modes in the nuclear wavefunction with momenta $k_\perp \leq Q_S$ have high occupation numbers typical of classical fields. In the CGC effective field theory, these classical fields are the dynamical degrees of freedom that couple stochastically to static light cone color sources $\rho^a({\bf x}_\perp)$ at large $x$~\cite{MV}; their source distribution is given by a gauge invariant weight functional $W_Y[\rho]$, where $Y= \ln(1/x)$ is the rapidity.  While this separation of degrees of freedom is arbitrary at some initial scale, its evolution with energy is described by the renormalization group (RG) equation ${\partial W_Y[\rho]\over \partial Y} = H_{\rm JIMWLK}[\rho]W_Y[\rho]$, where $H_{\rm JIMWLK}[\rho]$ is the JIMWLK Hamiltonian~\cite{JIMWLK}. The Balitsky-Kovchegov (BK) equation~\cite{BK} is a useful mean field (large $N_C$, large $A$) simplification of the JIMWLK equation describing, in closed form, the energy evolution of  the ``dipole" operator corresponding to the forward scattering amplitude in deep inelastic scattering. The JIMWLK and BK equations are derived in the leading logarithmic approximation in $x$, where running coupling effects are neglected. There is currently an intense on-going theoretical effort to compute next-to-leading order corrections to the kernels of the leading order RG equations~\cite{NLO}. 

\begin{figure}[htb]
\hfill
\includegraphics[width=0.4\textwidth]{shadcombowbcgcsncwide.eps}
\hfill
\includegraphics[width=0.4\textwidth]{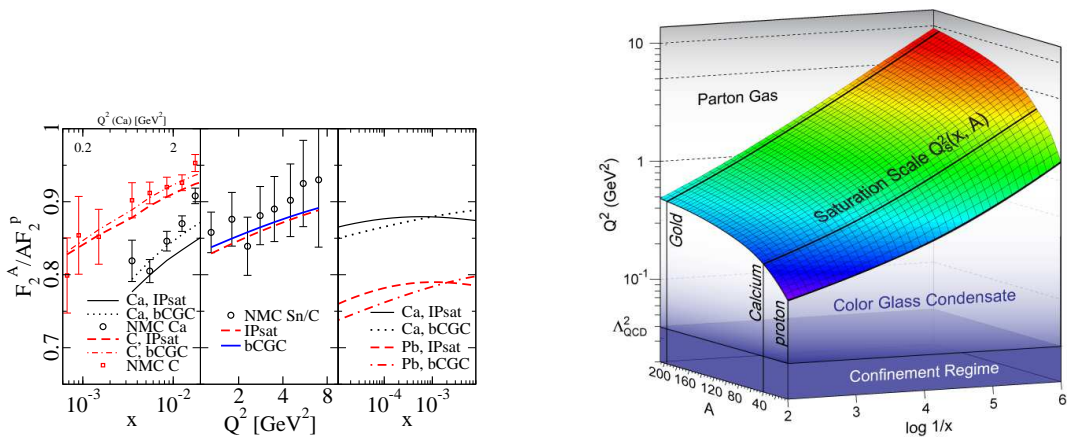}
\hfill
\caption{Left:Parameters fit to HERA inclusive data in the CGC motivated IPsat and bCGC models give excellent agreement with HERA 
diffractive and exclusive vector meson data. As shown, the same models, with no additional parameters, compared to fixed target NMC data; rightmost figure contains curves for 
fixed $Q^2=5$ GeV$^2$. See Ref.\cite{KLV} for further details. Right: The saturation scale for a quark probe (for gluons multiply $Q_S$ by color factor $\frac{9}{4}$) as a function of $x$ and $A$. The nuclear saturation scale extracted agrees with the value extracted from A+A multiplicity~\cite{Lappi0711} at RHIC to $\sim 10$\%.}
\label{fig:KLMV}
\end{figure}

Phenomenological ``dipole" models that incorporate the physics of saturation have been very successful-see Ref.~\cite{Raju-DIS} for  a recent review of comparisons of these models to the HERA, fixed target e+A, D+A and A+A RHIC data. A detailed dipole model study~\cite{KLV} shows that the saturation scale $Q_S^2(x,A) = A^{1/3}Q_S^2(x,A=1)$. See fig.~\ref{fig:KLMV} for partial results of this study. Because of the  nuclear ``oomph", saturation (CGC) effects should become increasingly visible in deuteron+gold collisions at RHIC,  p+A  collisions at the LHC, and especially at a future electron ion collider (EIC)~\cite{EIC}. 

\section{Multiparticle production in the Glasma at leading order}
\label{sect:three}
When two CGC sheets collide in an A+A collision, the static color sources in the nuclear wavefunctions become time dependent. An {\it ab initio} computation of the Glasma therefore requires a formalism to compute particle production in Quantum Field Theories coupled to external time dependent sources~\cite{GelisV,GelisLV}. Multi-particle production in the Glasma is always non--perturbative; the relevant question is whether the physics is one of strong coupling or weak coupling as will be considered here\footnote{The magnitude of the saturation scale is $Q_S^2 \sim 1$ -- $1.4$ GeV$^2$ at RHIC; estimates 
for the LHC are $Q_S^2\sim 2.6$ -- $3.9$ GeV$^2$~\cite{Lappi0711}.} 

For the inclusive gluon distribution, the leading order (LO) contribution is of order $O(\frac{1}{g^2})$ but all orders in $g\rho_{1,2}$. It can be expressed as 
\begin{eqnarray}
E_\p\frac{dN}{d^3\p}&=&\frac{1}{16\pi^3}
\lim_{x_0,y_0\to+\infty}\int d^3\x d^3\y
\;e^{ip\cdot(x-y)}
\;(\partial_x^0-iE_\p)(\partial_y^0+iE_\p)
\nonumber\\
&&\qquad\qquad\times\sum_{\lambda}
\epsilon_\lambda^\mu(\p)\epsilon_\lambda^\nu(\p)\;
\big<A_\mu(x)A_\nu(y)\big>\; .
\label{eq:AA}
\end{eqnarray}
The gauge fields on the right hand side can be computed numerically for proper times $\tau \geq 0$~\cite{KNV-Lappi} by solving the classical Yang-Mills (CYM) equations in the presence of the 
light cone current $J^{\mu,a} = \delta^{\mu +} \delta(x^-)\rho_1^a(x_\perp) +\delta^{\mu -}\delta(x^+) \rho_2^a(x_\perp)$ corresponding to 
the local color charge densities of the two nuclei with initial conditions determined by matching the (known) solutions to the CYM equations in the backward light cone. The energy densities of produced gluons can be computed in terms of $Q_S$ and one obtains $\varepsilon \sim 20$ -- $40$ GeV/fm$^3$ for the previously mentioned values of $Q_S$ obtained by extrapolating from fits to the HERA and fixed target e+A data. 

This LO formalism was applied to successfully predict the RHIC multiplicity at $y\sim 0$~\cite{KNV-Lappi} as well as the rapidity and centrality 
distribution of the multiplicities~\cite{KLN}.  CGC model comparisons to the RHIC data on limiting 
fragmentation~\cite{GSV} or solutions of CYM equations~\cite{Lappi0711}, extrapolated to the LHC, give $dN/d\eta|_{\eta\sim 0} \approx 1000$ -- $1400$ charged particles\footnote{See Ref.~\cite{Armesto-QM08} for other model predictions.}. At LO, the initial transverse energy is $E_T \sim Q_S$, which is 
about 3 times larger than the final measured $E_T$, while (assuming parton hadron duality) $N_{\rm CGC}\sim N_{\rm had.}$. The two conditions 
are consistent if one assumes nearly isentropic flow which reduces $E_T$ due to $PdV$ work while conserving entropy. This assumption has been 
implemented directly in hydrodynamic simulations~\cite{Hirano-Nara}. 

CGC based models give values for the initial eccentricity $\epsilon$ that are large than those in Glauber models~\cite{CGC-eccentricity} because the energy and number density locally is  sensitive to the lower of the two saturation scales (or local participant density) in the former and the average of the two in the latter. Naively, CGC initial conditions would have more flow then and have more room for dissipative effects relative to Glauber. This conclusion is turned on its head in a simple parametrization of incompletely thermalized flow~\cite{BBBO}: 
$v_2/\epsilon = {(v_2/\epsilon)_{\rm hydro} \over (1+ K/K_0)}$, where $K = \frac{1}{S_\perp}\sigma \frac{dN}{dy} c_s$ is the Knudsen number, $\sigma$ the 
cross-section, $c_s$ the sound speed and $S_\perp$ the transverse overlap area. $K_0$ is a number of order unity. If thermalization were complete, 
$K\rightarrow 0$ and one approaches the hydro bound. Computing $\epsilon$ with different initial conditions, and plotting the l.h.s ratio versus $\frac{1}{S_\perp}\frac{dN}{dy}$, one has a two parameter fit to $\sigma$ and $c_s$. The greater CGC eccentricity forces $v_2/\epsilon$ to be lower for more central collisions thereby leading to lower 
$c_s$;  quicker saturation of $v_2/\epsilon$ forces larger $\sigma$ and therefore lower $\eta$ in the CGC relative to Glauber~\cite{DO}. It is conceivable however that this result may not prove robust against more detailed modeling. Nevertheless, it is clear that the results are very sensitive to the initial conditions. 

How much flow is generated in the Glasma before thermalization? The primordial Glasma has 
occupation numbers $f\sim \frac{1}{\alpha_S}$ and can be described as a classical field. As the Glasma expands, higher momentum modes increasingly 
become particle like and eventually the modes have occupation numbers $f<1$, which may be described by a thermal spectrum.  A first computation of elliptic flow of the Glasma found only about half the observed elliptic flow~\cite{KV-PLB} albeit the computation did not properly treat the interaction between hard and soft modes in the Glasma. Formulating a kinetic theory that describes this evolution is a challenging problem in heavy ion collisions--for a preliminary discussion, see ~\cite{Jeon}.

The LO Glasma result, from the solutions of Yang--Mills equations,  has very interesting properties. Firstly, the 
solution is boost invariant in the strong sense--the fields are independent of the space-time rapidity; the dynamics of the produced gluon fields is purely transverse as a function of proper time. Another interesting feature is that the chromo $E$ and $B$ fields are purely longitudinal after the collision~\cite{Glasma}. This result suggests that one can generate topological Chern--Simons charge in heavy ion collisions~\cite{glasma-sphaleron}. Because the range of color correlations in the transverse plane is of order $1/Q_S$, the LO picture that emerges is of color flux tubes with finite topological charge, stretching between the valence color degrees of freedom. As we shall see, this picture provides a plausible explanation of the near side ridge.

\begin{figure}[htb]
\hfill
\includegraphics[width=0.25\textwidth]{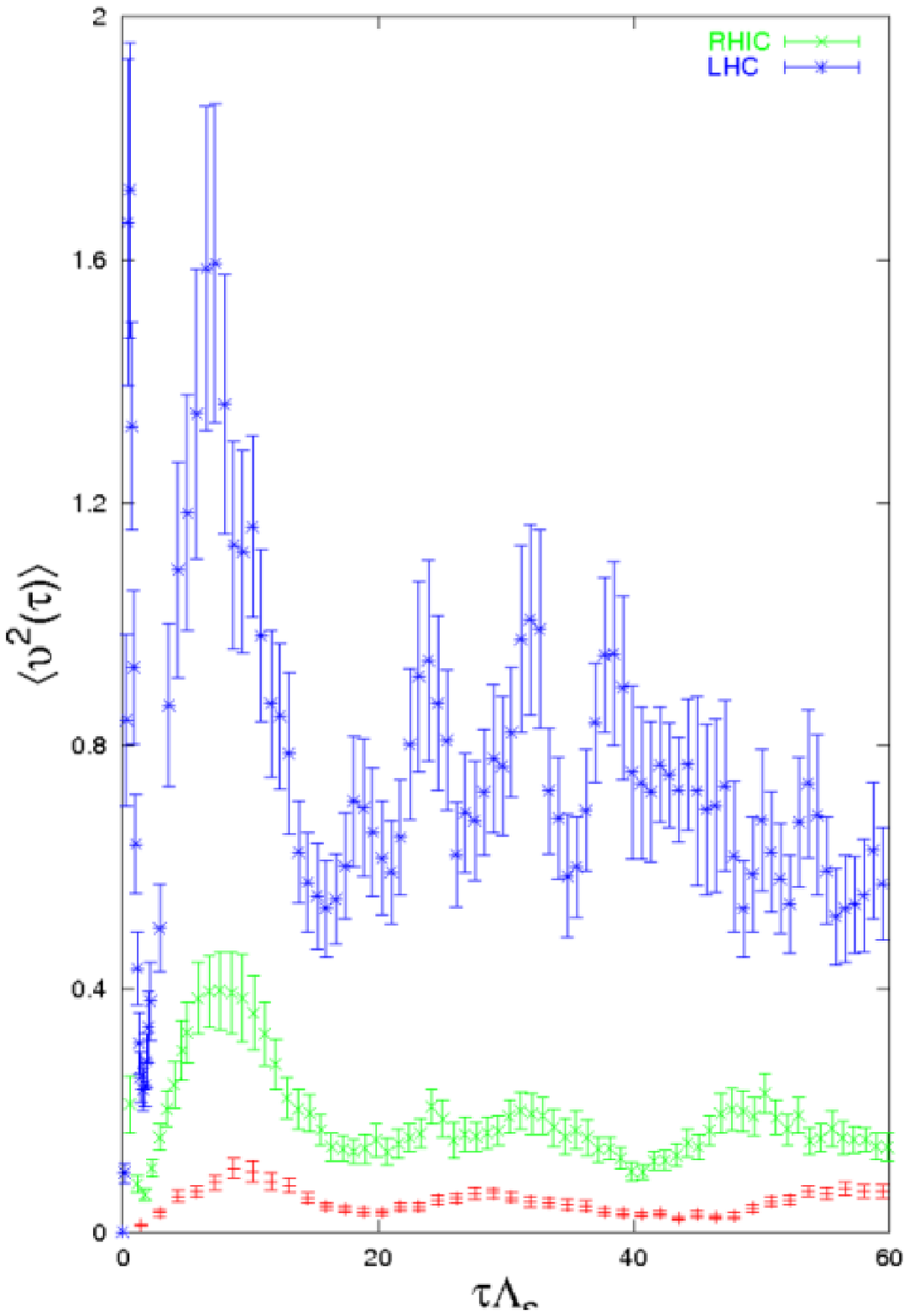}
\hfill
\includegraphics[width=0.5\textwidth]{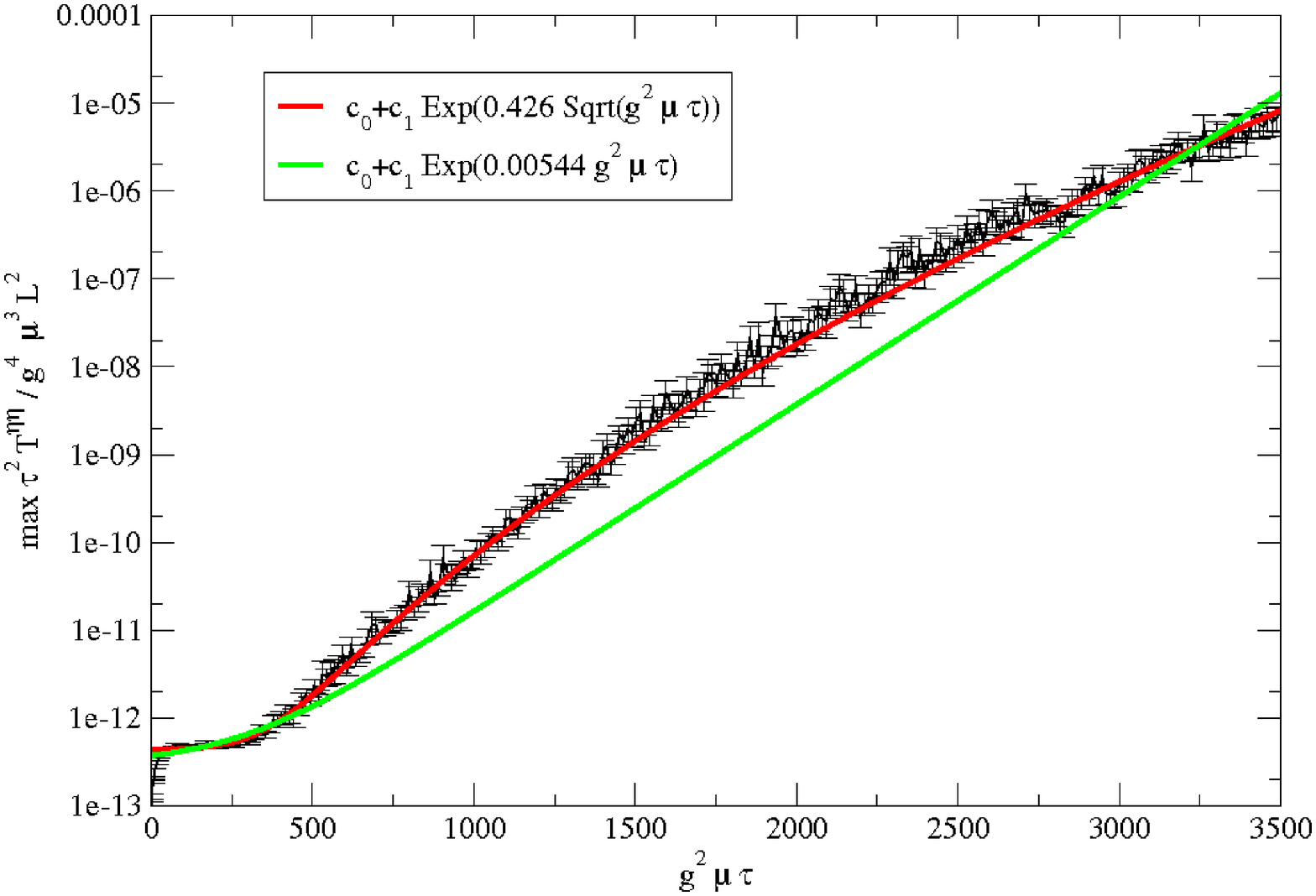}
\hfill
\caption{Left: Chern--Simons mean squared charge as a function of proper time (in units of the saturation scale), generated in the LO boost-invariant 2+1-D Glasma~\cite{glasma-sphaleron}. Quantum fluctuations allow sphaleron transitions which may give rise to significantly greater values of $\sqrt{\langle \nu^2\rangle}$. Right: The 
same quantum fluctuations, albeit suppressed by $\alpha_S$, grow as $\alpha_S\exp(\sqrt{Q_S\tau})$~\cite{PaulR}. These unstable fluctuations, 
resummed to all orders, likely give rise to more isotropic initial distributions in the Glasma~\cite{GelisLV}.}
\label{fig:LO}
\end{figure}

The LO field configurations are very unstable and lead to very anisotropic momentum distributions at later times. Such distributions can trigger an instability analogous to the Weibel instability in QED plasmas~\cite{Stan}. Romatschke and I showed (in 3+1-D numerical solutions of CYM equations)~\cite{PaulR} that small rapidity dependent quantum fluctuations in the initial conditions generate transverse $E$ and $B$ fields  that grow rapidly as $\exp(\sqrt{Q_S\tau})$. They are the same size as the rapidly diluting longitudinal $E$ and $B$ fields on time scales of order $\frac{1}{Q_S}\ln^2(\frac{C}{\alpha_S})$. The transverse $E$ and $B$ fields may cause large angle deflections of colored particles leading to $p_T$ broadening and energy loss of jets--numerical simulations by the Frankfurt group appear to confirm this picture~\cite{Frankfurt}. These interactions of colored high momentum particle like modes with the soft coherent classical field modes may also generate a small ``anomalous viscosity" whose effects on transport in the Glasma may 
mask a larger kinetic viscosity~\cite{Duke}. The same underlying physics may cause ``turbulent isotropization" by rapidly transferring momentum from soft ``infrared" longitudinal modes to ultraviolet modes~\cite{turbulence}. Finally, albeit the LO result demonstrated that one could have non-trivial Chern-Simons charge in heavy ion collisions, the boost invariance of CYM equations disallows sphaleron transitions that permit large changes in the Chern-Simons number~\cite{glasma-sphaleron}. With rapidity dependent quantum fluctuations, sphaleron transitions can go. These may have important consequences--in particular $P$ and $CP$ odd metastable transitions that cause a novel ``Chiral Magnetic Effect"~\cite{Warringa} in heavy ion collisions. 
Numerical CYM results for Chern-Simons charge and (square root) exponential growth of instabilities are shown in fig.~\ref{fig:LO}.

\section{QCD Factorization and the Glasma instability}
\label{sect:four}
\begin{figure}[htb]
\hfill
\includegraphics[width=0.5\textwidth]{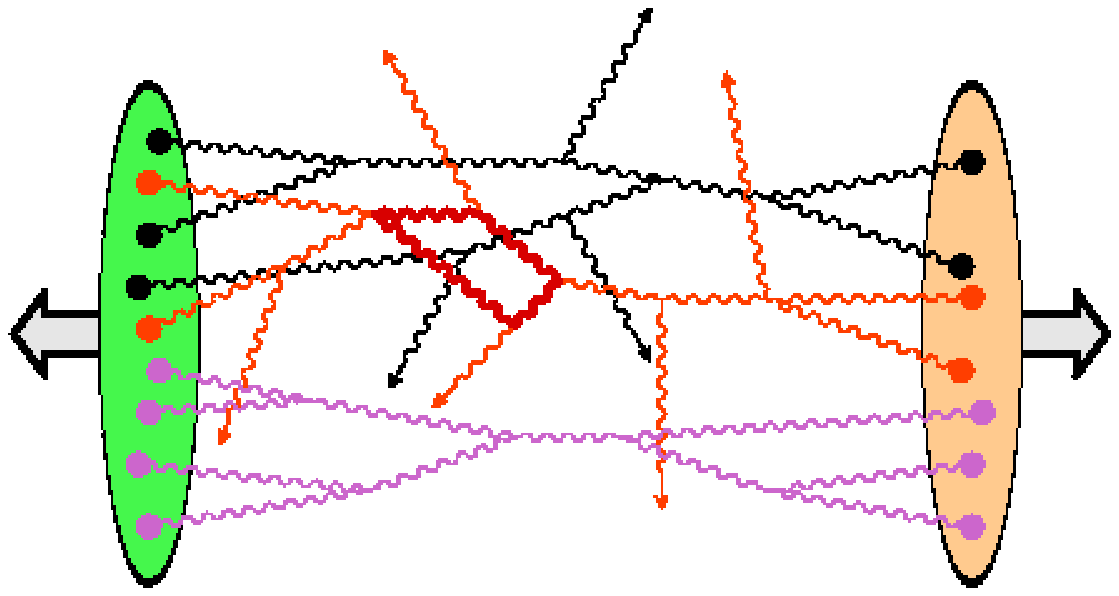}
\hfill
\includegraphics[width=0.2\textwidth]{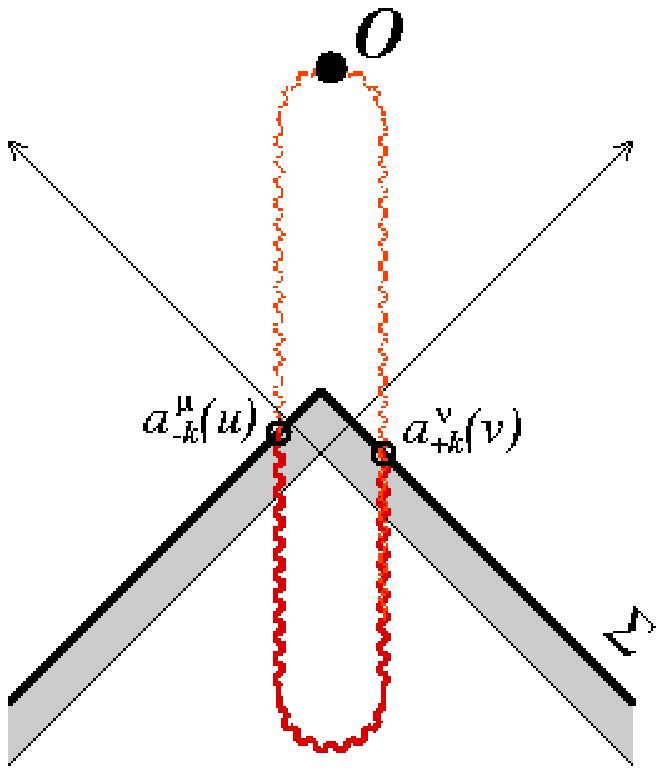}
\hfill
\caption{Left: Illustration of particle production in a field theory with strong time dependent sources. Right: Quantum fluctuations, where the 
nuclei talk to each other before the collision, are suppressed; this is responsible for high energy factorization of inclusive gluon operators ${\cal O}$ in A+A collisions~\cite{GelisLV}.}
\label{fig:fact}
\end{figure}

The discussion at the end of the last section strongly suggests that next-to-leading order (NLO) quantum fluctuations in the Glasma, while (superficially) parametrically suppressed, may alter our understanding of heavy ion collisions in a fundamental way. To cosmologists, this will not come as a surprise--quantum fluctuations play a central role there as well. In recent papers, it was shown for a scalar theory that moments of the multiplicity distribution at NLO in A+A collisions could be computed as an initial value problem with retarded boundary conditions; this framework has now been extended to QCD~\cite{GelisV}. In QCD, the problem is subtle because some quantum fluctuations occur in the nuclear wavefunctions and are responsible for how the wavefunctions evolve with energy; others contribute to particle production at NLO. Fig.~\ref{fig:fact} illustrates particle production in fields theories with 
strong sources and the non-factorizable quantum fluctuations  that are suppressed in the leading log framework.  

A factorization theorem organizing these quantum fluctuations shows that all order leading logarithmic contributions to an inclusive gluon operator ${\cal O}$ in the Glasma gives~\cite{GelisLV}
\begin{equation}
\langle {\cal O}\rangle_{_{\rm LLog}}
= \int [D{\wt{\rho}}_1] [D{\wt{\rho}}_2]\;
W_{Y_{\rm beam}-Y}[\wt{\rho}_1]\, W_{Y_{\rm beam}+Y}[\wt{\rho}_2]\;
{\cal O}_{_{\rm LO}}\left[\wt{\rho}_1,\wt{\rho}_2\right] \; ,
\label{eq:fact-formula}
\end{equation}
where ${\cal O}_{_{\rm LO}}$ is the same operator evaluated at LO by solving classical Yang--Mills equations and $W_{Y_{\rm beam}\mp Y}[\wt{\rho}_{1,2}]$ are the weight functionals (introduced in section~\ref{sect:two}) that obey the JIMWLK Hamiltonians describing the rapidity evolution of the projectile and target wavefunctions respectively. This theorem is valid if the rapidity interval corresponding to the production of the final state, $\Delta Y \leq \frac{1}{\alpha_S}$. The $W$'s are analogous to the parton distribution functions in collinear factorization; determined non-perturbatively at some initial scale $Y_0$, their evolution with $Y$ is given by the JIMWLK Hamiltonian. 

This factorization theorem  is a necessary first step before a full NLO computation of gluon production in the
Glasma. Eq.~(\ref{eq:fact-formula}) includes only the NLO terms that are enhanced by a large logarithm of $1/x_{1,2}$, while the complete NLO calculation would also include non enhanced terms. These would be
of the same order in $\alpha_S$ as the production of quark-antiquark pairs \cite{GelisKL} from the classical field. To be really useful, this complete NLO calculation likely has to be promoted to a Next-to-Leading Log (NLL) result by resumming all the terms in $\alpha_S(\alpha_S\ln(1/x_{1,2})^n$. Now that evolution equations
in the dense regime are becoming available at NLO, work in this direction is a promising prospect~\cite{NLO}. 

In addressing the role of instabilities at NLO, note that small field fluctuations fall into three classes:
i) Zero modes ($p_\eta=0$) that generate the leading logs  resummed in eq.~\ref{eq:fact-formula}; the coefficients of the leading logs do not depend on $x^\pm$. ii) Zero modes that do not contribute at leading log because they are less singular than the leading log contributions. These become relevant in resumming NLL corrections to the factorization result~\cite{NLO}. Because they are zero modes, they do not trigger plasma instabilities. iii) Non zero modes ($p_\eta\not=0$) that  do not contribute  large logarithms of $1/x_{1,2}$, but grow exponentially  as $\exp(\sqrt{Q_S\tau})$.  While these boost non-invariant terms are suppressed by $\alpha_S$, they are enhanced by exponentials of the proper time after the collision. These leading temporal divergences can be resummed  and the expression for inclusive gluon operators in the Glasma revised to read 
\begin{eqnarray}
&&
\left<{\cal O}\right>_{\rm LLog+LInst}
=
\int [D\wt{\rho}_1] [D\wt{\rho}_2]\;
W_{Y_{\rm beam}-Y}[\wt{\rho}_1]\, W_{Y_{\rm beam}+Y}[\wt{\rho}_2]
\nonumber\\
&&\qquad\qquad\qquad\times
\int\big[Da(\vec\u)\big]\;\widetilde{Z}[a(\vec\u)]\;
{\cal O}_{_{\rm LO}}[\wt{\cal A}^+_1+a,\wt{\cal A}^-_2+a]
\label{eq:final}
\end{eqnarray}
where $\wt{\cal A}_{1}^+(x)=-\frac{1}{{\bs\partial}_\perp^2}\,\wt{\rho}_{1}(x_\perp,x^-)$ and $\wt{\cal A}_{2}^-(x)=-\frac{1}{{\bs\partial}_\perp^2}\,\wt{\rho}_{2}(x_\perp,x^+)$. The effect of the resummation of instabilities is therefore  to add fluctuations to the
initial conditions of the classical field, with a distribution that depends on the outcome of the resummation. This spectrum 
$\widetilde{Z}[a(\vec\u)]$ is the final incomplete step in determining all the leading singular contributions to particle production in the Glasma-see however Ref.\cite{FukusGM1}. The stress-energy tensor $T^{\mu\nu}$ can then be determined {\it ab initio} and matched smoothly to kinetic theory or hydrodynamics at late times. 

\section{Two particle correlations in the Glasma and the Ridge}

Striking ``ridge'' events were revealed in studies of the near
side spectrum of correlated pairs of hadrons at RHIC~\cite{Ridge-expt}. The spectrum of correlated pairs on
the near side of the STAR detector extends across the entire detector acceptance in
pseudo-rapidity of order $\Delta \eta\sim 2$ units but is strongly
collimated for azimuthal angles $\Delta \phi$.  Preliminary analyses
of measurements by the PHENIX and PHOBOS collaborations corroborate the STAR results.  In the latter
case, the ridge is observed to span the PHOBOS acceptance in pseudo-rapidity of $\Delta \eta \sim 3-4$ units.  

\begin{figure}[htb]
\hfill
\includegraphics[width=0.25\textwidth]{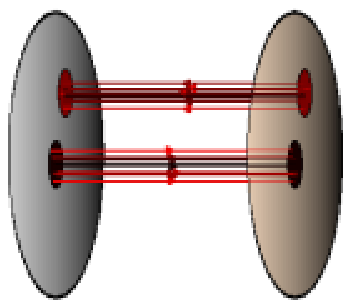}
\hfill
\includegraphics[width=0.45\textwidth]{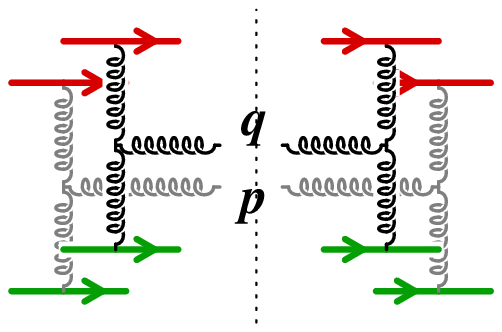}
\hfill
\caption{Left:Glasma flux tubes of transverse size $\frac{1}{Q_S} < \frac{1}{\Lambda_{\rm QCD}}$. The field lines correspond to parallel $E$ and 
$B$ fields, which carry topological charge. Right: The leading two particle contribution. Superficially disconnected, they are connected by 
averaging over the large $x$ color sources. Systematic power counting shows these graphs dominate over usual ``pQCD" graphs at high energies.}
\label{fig:Ridge}
\end{figure}

Causality dictates (in strong analogy to CMB superhorizon fluctuations) that long range rapidity correlations causing the ridge must have occured at proper times $\tau\le \tau_{\rm freeze\ out}\;e^{-\frac{1}{2}|y_{_A}-y_{_B}|}$, 
where $y_{_A}$ and $y_{_B}$ are the rapidities of the correlated particles. If the ridge span in psuedo-rapidity is large, these correlations must have originated in the Glasma. As noted previously in section~\ref{sect:three}, particles  produced from Glasma 
flux tubes are boost invariant. See fig.~\ref{fig:Ridge}. Correlated two gluon production in the Glasma flux tube can be shown to be independent of rapidity~\cite{DumitruGMV}. Ours is the only dynamical model with this feature-for other models, see Ref.\cite{Hwa}.  The particles produced in 
a flux tube are isotropic locally in the rest frame but are collimated in azimuthal angle when boosted by transverse flow~\cite{Shuryak-Voloshin}. Combining our dynamical calculation of two particle correlations with a simple ``blast wave" model of transverse flow, we obtain reasonable agreement with 200 GeV STAR data on the amplitude of the correlated two particle spectrum relative to the number of binary collisions per participant pair. This simple model also gives reasonable agreement for the amplitude for collisions at 62 GeV~\cite{unpublished}. One caveat is that the collimation in azimuthal angle is weaker than seen in the data-this can be significantly improved with a more refined treatment of transverse flow. The Glasma flux tube model has several additional attractive features consistent with observations-see Ref.~\cite{DumitruGMV} for a fuller discussion.

\section*{Acknowledgments}
I  thank A. Dumitru, S. Gavin,T. Lappi, L. McLerran, P. Sorensen and especially F. Gelis for useful discussions . This manuscript was authored under DOE Contract No.~\#DE-AC02-98CH10886.

\end{document}